\documentclass{article}
\usepackage{spconf,amsmath,amssymb,graphicx}
\usepackage{url}

\title{SPEECH-BASED EMOTION RECOGNITION WITH SELF-SUPERVISED MODELS USING ATTENTIVE CHANNEL-WISE CORRELATIONS AND LABEL SMOOTHING}
\name{Sofoklis Kakouros$^{1}$, Themos Stafylakis$^{2}$, Ladislav Mo\v{s}ner$^{3}$, {Luká\v{s} Burget$^{3}$}
}
\address{$^{1}$University of Helsinki, Finland\\ 
$^{2}$Omilia - Conversational Intelligence, Athens, Greece\\
$^{3}$Brno University of Technology, Faculty of Information Technology, Speech@FIT, Czechia}
\begin{document}
%\ninept
%
\maketitle
\begin{abstract}
When recognizing emotions from speech, we encounter two common problems: how to optimally capture emotion-relevant information from the speech signal and how to best quantify or categorize the noisy subjective emotion labels. Self-supervised pre-trained representations can robustly capture information from speech enabling state-of-the-art results in many downstream tasks including emotion recognition. However, better ways of aggregating the information across time need to be considered as the relevant emotion information is likely to appear piecewise and not uniformly across the signal. For the labels, we need to take into account that there is a substantial degree of noise that comes from the subjective human annotations. In this paper, we propose a novel approach to attentive pooling based on correlations between the representations' coefficients combined with label smoothing, a method aiming to reduce the confidence of the classifier on the training labels. We evaluate our proposed approach on the benchmark dataset IEMOCAP, and demonstrate high performance surpassing that in the literature. The code to reproduce the results is available at \url{github.com/skakouros/s3prl_attentive_correlation}.
\end{abstract}
\section{INTRODUCTION}
\label{sec:intro}

Emotional expressions are a fundamental component of spoken interaction. When we communicate with other people, we are implicitly monitoring their emotional state and respond based on that emotional state \cite{brave2007emotion}. Emotions fall in the realm of prosodic function. In recent years, the importance of prosodic qualities in speech has attracted increasing attention. This is also the case for speech emotion recognition (SER) which has seen a growing interest with the increasing role of spoken language interfaces in human-computer interaction (HCI) applications \cite{lee2005toward}. However, recognizing emotions in speech remains a challenging problem complicated by numerous factors including fundamental issues of how emotion is defined, elicited, expressed, and communicated \cite{picard2000affective} and extending to how we can capture this information from speech.

The challenges in SER can be split into three distinct problems. First, we encounter the issue of developing engineered representations that can robustly capture the acoustic information in speech that best describe the variation found across different emotions. This has been traditionally done using features such as mel-frequency cepstral coefficients (MFCCs), filterbanks, fundamental frequency, energy, zero-crossing rate, chroma-based features and their feature functionals \cite{ ververidis2006emotional} or through standardized feature collections and their functionals such as eGeMAPS \cite{pepino2021emotion}. More recently, self-supervised learning (SSL) has shown its effectiveness in various domains, including SER, and is becoming the new principle for extracting representations from speech. HuBERT \cite{hsu2021hubert}, Wav2vec 2.0 \cite{baevski2020wav2vec}, WavLM \cite{chen2021wavlm}, are some of the self-supervised approaches for speech representation learning that have been used in the context of SER \cite{ yang2021superb,pepino2021emotion}.

The second issue that we face in SER is the effective modeling of the long temporal context over which emotions take place. Emotion specific information lies beyond segmental productions and in longer time scales. These may include parts of an utterance but can also span across one or more utterances. To appropriate model long-term dependencies by capturing and connecting the relevant cues across time suitable methods are necessary. These vary from approaches that simply take the first and second order statistics of self-supervised representations across time \cite{yang2021superb} to approaches that focus on complex sequence modeling tasks \cite{sarma2018emotion}. For example Sarma et al. \cite{sarma2018emotion} used a TDNN architecture combined with LSTM and self-attention to model the long-term temporal context and to capture the emotionally relevant portions of speech. In a recent work, Liu et al. \cite{liu2022discriminative} used a cascaded attention network to locate the relevant emotion regions from the input features. Other approaches also use different types of recurrent neural networks (RNNs) to explain the long temporal contexts of emotions in speech \cite{lee2015high}.

The third and final issue comes from the observation that human emotional expressions are often unclear and ambiguous, leading to disagreement and confusion among human evaluators \cite{kim2015leveraging,mower2009interpreting}. This confusion might be partly attributed to the multimodal nature of emotion expression. Facial expressions, hand gestures, and speech with its prosodic and linguistic content all work together in eliciting different emotions. Perhaps the absence of multimodality may be one source of confusion that leads to overlaps in the clusters of the different emotion classes. However, speech alone holds much relevant information in its prosodic content that can be used for robust SER. Different ways to tackle the problem of noisy labels and consequently the uncertainty in predicting emotions have been suggested in the literature. These typically include custom loss functions \cite{liu2021speech,liu2022discriminative} and modifications to the target hard labels \cite{fayek2016modeling,tarantino2019self}. For example Liu and Wang used a triplet loss to make anchor utterances more similar to all other positive utterances \cite{liu2021speech} while Tarantino et al. used regression targets instead of hard categorical targets by taking the proportion of the classes within the annotations \cite{tarantino2019self}.

This paper presents a framework for SER that uses pre-trained speech models with a novel approach to attentive pooling based on channel-wise correlations on soft targets. We evaluate the framework with HuBERT \cite{hsu2021hubert}, Wav2vec 2.0 \cite{baevski2020wav2vec}, and WavLM \cite{chen2021wavlm} upstream models. We use the SUPERB \cite{yang2021superb} evaluation setup throughout our experiments. The effectiveness of our proposed framework is evaluated on the interactive emotional dyadic motion capture (IEMOCAP) dataset \cite{busso2008iemocap} and shows state-of-the-art performance in SER.

\section{RELATION TO PRIOR WORK}
\label{sec:relatedwork}

The idea of taking the correlations between different filter responses over the spatial extent of the feature maps to obtain a representation of the style of an input image was introduced by Gatys et al. \cite{gatys2015neural}. Their method was later adapted to speech where it has found applications in speech generation and voice conversion \cite{chorowski2018using}, pooling to obtain speaker embeddings \cite{stafylakis2021speaker}, and sentence-level tasks such as speaker identification, speaker verification, and SER \cite{stafylakis2022corrpooling}.

In this work, we extend the method for correlation pooling presented in \cite{stafylakis2022corrpooling}. In \cite{stafylakis2022corrpooling}, it was shown that channel-wise correlations provide an alternative way of extracting speaker and emotion information from self-supervised models, providing also improvements over the standard mean and mean-std pooling (std stands for standard deviation). In the present work, we add an attention mechanism to pool representations before estimating the correlation matrix, while reducing the confidence on the target labels with label smoothing.

\section{PROPOSED METHOD}
\label{sec:proposedmethod}

\begin{figure}[tb] %htb
\begin{minipage}[b]{1.0\linewidth}
  \centering
  \centerline{\includegraphics[width=\linewidth]{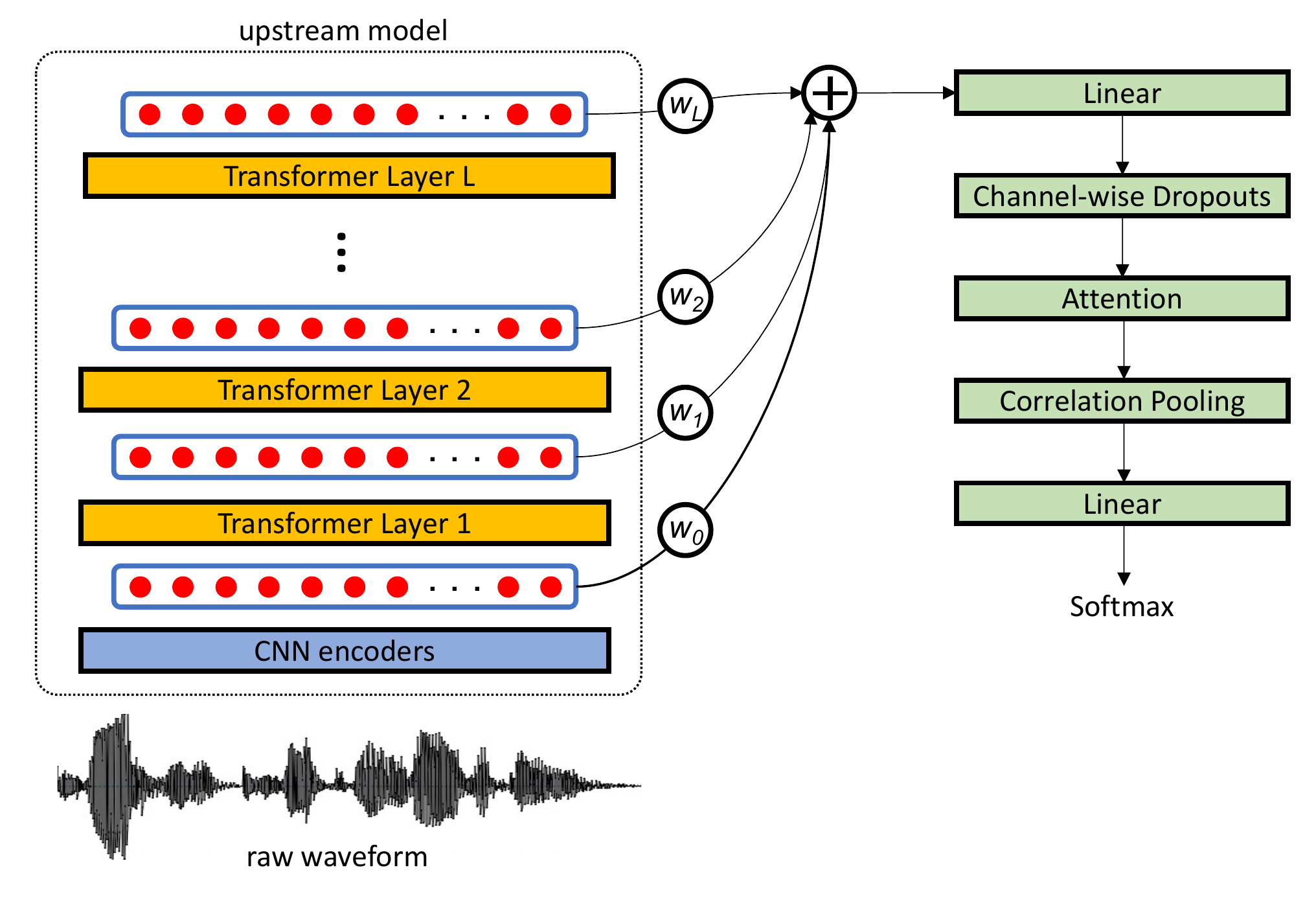}}
\end{minipage}
\caption{Overview of the proposed architecture.}
\label{fig:modelsetup}
\end{figure}

We construct our SER framework based on the pipeline and principles of SUPERB \cite{yang2021superb}. As finetuning pretrained models has a high resource demand in terms of the computational power needed, we use a simple framework with a frozen pretrained model and lightweight classification heads. An overview of our framework is presented in Fig. \ref{fig:modelsetup}. In this section we describe details of the proposed approach.

\subsection{Layer-wise pooling}
\label{ssec:layerpooling}

We extract information relevant for our downstream task as in SUPERB (\cite{yang2021superb}), by taking representations from all transformer layers in the model and collapsing them to one via a weighted average (see Fig. \ref{fig:modelsetup}). There is one weight for each layer (a total of $L+1$) and all weights are trained jointly with the classification network. The weighted average representation is expressed as follows

\begin{equation}
\label{eq:layerwise}
{\bf h}_t = \sum_{l=0}^{L}\gamma_{l}{\bf h}_{t,l}, 
\end{equation}
where the weights $\sum_{l=0}^{L}\gamma_{l}=1, \, \gamma_{l} \geq 0$ 
are implemented with a learnable vector of size $L+1$, followed by the Softmax function, and ${\bf h}_{t,l}$ is the representation of the $l$th layer at time $t$ (${\bf h}_{t,0}$ is the output of the ConvNet). 

\subsection{Frame-wise pooling}
\label{ssec:meanpooling}
Tasks requiring sentence-level classification typically employ a pooling method, such as mean, max or attentive pooling. Mean pooling, which is employed in SUPERB is defined as
\begin{equation}
    {\bf r} = \bar{\bf h} = \frac{1}{T}\sum_{t=1}^{T}{\bf h}_{t}, 
\end{equation}
where $T$ is the number of acoustic features of an utterance extracted by the ConvNet, ${\bf r}$ is the resulting pooled representation, while ${\bf h}_{t}$ are the representations at time $t$ after layer-wise pooling. Concatenating the pooled representations with std features is in general helpful in speaker recognition \cite{wang2021pooling}, and is implemented as
\begin{equation}
    {\bf r} = \left[\bar{\bf h} ; \left(\frac{1}{T}\sum_{t=1}^{T}({\bf h}_{t} - \bar{\bf h})^2 \right)^{1/2}\right], 
\end{equation}
where $[\cdot;\cdot]$ denotes vector concatenation and the exponents should be considered as element-wise operators.

\subsection{Correlation pooling}
\label{ssec:correlationpooling}
Correlation pooling (introduced in \cite{stafylakis2021speaker}) is an alternative pooling method which has shown improvements in speaker recognition. The embedding dimension of SSL models (typically $768$ or $1024$) is too high to estimate correlations. We therefore project ${\bf h}$ onto a lower $d_v$-dimensional space ${\bf v}$ via a linear layer ($d_v = 256$). We then calculate the mean vector ${\boldsymbol \mu}$ and the covariance matrix ${\bf \Sigma}$ of ${\bf v}$ as follows
\begin{equation}
\label{eq:cov_pooling}
    {\boldsymbol \mu} = \frac{1}{T} \sum_{t=1}^T {\bf v}_t, \, \,
    {\bf \Sigma} = \frac{1}{T}\sum_{t=1}^{T}({\bf v}_{t} - {\boldsymbol \mu} )({\bf v}_{t} - {\boldsymbol \mu} )', 
\end{equation}
where ${\bf x}'$ denotes the transpose of ${\bf x}$. Finally, the correlation matrix is derived by normalizing with respect to the variances as follows 
\begin{equation}
\label{eq:corr_pooling}
{\bf C} = {\bf \Sigma} \oslash {\bf S},
\end{equation}
where ${\bf S} = {\bf s} {\bf s}' + \epsilon {\bf 1}$, ${\bf s} = \mbox{diag}({\bf \Sigma})^{1/2}$ (i.e. the vector of std), $\oslash$ denotes element-wise division, while $\epsilon=10^{-8}$.
Since ${\bf C}$ is a symmetric matrix and its diagonal elements are equal to 1, we vectorize the elements above the diagonal, yielding a $(d_v\times(d_v-1)/2)$-sized vector, which we project onto a linear layer followed by the Softmax over the emotion classes. For regularization, dropout is applied to ${\bf v}$, where whole channels are dropped with probability $p_d = 0.25$. 

\subsection{Attentive correlation pooling}
\label{ssec:layerpooling}
We introduce here the attentive correlation pooling, by inserting weights in the estimates of the statistics, i.e.   
\begin{equation}
{\boldsymbol \mu} = \sum_{t=1}^T w_t{\bf v}_t, \, \, {\bf \Sigma} = \sum_{t=1}^{T}w_t({\bf v}_{t} - {\boldsymbol \mu} )({\bf v}_{t} - {\boldsymbol \mu} )' 
\end{equation}
%and
%\begin{equation}
%\label{eq:cov_att_pooling}
%    {\bf \Sigma} = \sum_{t=1}^{T}w_t({\bf v}_{t} - {\boldsymbol \mu} )({\bf v}_{t} - {\boldsymbol \mu} )'. 
%\end{equation}
and we calculate ${\bf C}$ as in eq. (\ref{eq:corr_pooling}).
The weights $\{w_t\}_{t=1}^T$ (where $\sum_t w_t=1$ and $w_t \geq 0$) are estimated using a new flavor of attention. Similarly to the single-head attention a single set of weights is estimated. Similarly to the multi-head attention, multiple heads are employed, however their similarities with ${\bf v}_t$ are aggregated prior to the Softmax function via the log-sum-exp function, as follows
\begin{equation}
    \{w_t\}_{t=1}^{T} = \mbox{Softmax}\left(\{a_t\}_{t=1}^{T}\right), 
\end{equation}
where
\begin{equation}
    a_t = \log \sum_h \exp\left({\bf q}_h'{\bf o}_t + b_h\right), 
\end{equation}
the heads $\{{\bf q}_h,b_h\}_{h=1}^H$ are trainable $d_v$-dimensional vectors and biases, ${\bf o}_t = \mbox{ReLU}({\bf W}_{att}{\bf v}_t)$ and ${\bf W}_{att}$ is a square matrix ($d_v\times d_v$). Note that an equivalent implementation is to use as input to the Softmax the $(H\times T)$-sized \emph{vector} of dot-products ${\bf q}_h'{\bf o}_t + b_h$ and sum the outputs over heads to obtain $\{w_t\}_{t=1}^{T}$. 

As we observe, the proposed attention resembles a mixture model with heads parametrizing the mixture components. The log-sum-exp function is a soft version of the max operator, meaning that $a_t$ is high when at least one of the $H$ head-specific dot-products $\{ {\bf q}_h'{\bf o}_t +b_h\}_{h=1}^H$ is high.

The rationale for proposing this kind of attention is two-hold. We desire to keep the multi-modality of multi-head attention since a single head is too weak to capture the phonetic, speaker, emotion and channel variability. On the other hand, the standard multi-head attention results in $H$ context-vectors (in our case $H$ correlation matrices), which can be hard to estimate robustly, especially when the utterances are short and the estimation involves second-order statistics.    

\subsection{Label smoothing}
\label{ssec:layerpooling}

%The cross-entropy loss typically employed for classification is defined as follows
%\begin{equation} 
%\label{eq:crossentropy}
%H({\bf y}, {\bf p}) = - {\bf y}' \log{\bf p} = -\sum_{k=1}^K y_k\log p_k
%\end{equation}
%where ${\bf y}$ is the one-hot vectorial representation of the target class, and ${\bf p}$ is the vector of  posterior probabilities $\{p_k\}_{k=1}^K$ that the model assigns to each of the $K$ classes. 
With label smoothing we soften the hard (one-hot) targets vectors ${\bf y}$ of the training set as follows
\begin{equation}
\label{eq:labelsmoothing}
{\bf y}^{LS} = {\bf y}(1-p_l) + p_l/K, 
\end{equation}
where $p_l$ is the label smoothing parameter (i.e. the probability mass equally distributed to all classes) and ${\bf y}^{LS}$ is the smoothed target vector \cite{szegedy2016rethinking}. Cross-entropy is still employed as loss function, but with soft targets.

\section{EXPERIMENTS}
\label{sec:experiments}

\subsection{Datasets}
\label{ssec:datasets}
The IEMOCAP database consists of multi-modal recordings (speech, video) by 10 actors in dyadic sessions in English ($\approx 12$ hours) \cite{busso2008iemocap}. The dataset is split in 5 dialogue sessions (one female-male speaker pair per session). The emotions conveyed are happiness, anger, excitement, sadness, surprise, fear, frustration, and neutral state. As in other studies on IEMOCAP, we relabel excitement as happiness and use 4 balanced emotion classes, namely: anger, happiness, sadness, and neutral \cite{yang2021superb, pepino2021emotion, fayek2017evaluating}. All other classes are discarded.

%The Ryerson Audio-Visual Database of Emotional Speech and Song (RAVDESS) is a multimodal database consisting of recordings from 24 professional actors (12 female) in North American accent \cite{ livingstone2018ryerson}. All actors enact the statements “Kids are talking by the door” and “Dogs are sitting by the door” with different emotions, at two intensity levels (normal and strong), and in speech and song conditions. Speech recordings were expressed with the emotions of calm, happy, sad, angry, fearful, surprise, disgust, and neutral. Song recordings were elicited with the emotions of calm, happy, sad, angry, fearful, and neutral. The dataset contains a total of 7356 recordings. Each recording was rated 10 times on emotional validity, intensity, and genuineness by 247 individuals. In our experiments we used only the speech portion of the corpus that contains 1440 audio files recorded by 12 male and 12 female actors. As there is no standard split available for the corpus, we opted for a speaker independent evaluation that leaves approximately 16\% of the recordings in a held-out set. Thus, we kept speakers 21 (male), 22 (female), 23 (male), 24 (female) in the test set and used the remaining speakers for training and validation. Following \cite{ pepino2021emotion, venkataramanan2019emotion} we merged the neutral and calm classes, resulting into a total of 7 emotions. %

\subsection{Experimental Setup}
\label{ssec:experimentalsetup}
We use a 5-fold cross-validation setup where at each fold we leave out one session from the dataset. Each held-out session consists of two speakers that are not present in the train and validation sets. This approach leaves approximately 19\% of the data for testing. Mean and standard deviation across folds is computed and presented as the aggregated result. Our SER framework is evaluated with WavLM, Wav2vec 2.0, and HuBERT speech representations.

\section{RESULTS AND ANALYSIS}
\label{sec:results}

\begin{figure}[b]
\begin{minipage}[b]{0.48\linewidth}
  \centering
  \centerline{\includegraphics[width=4.6cm]{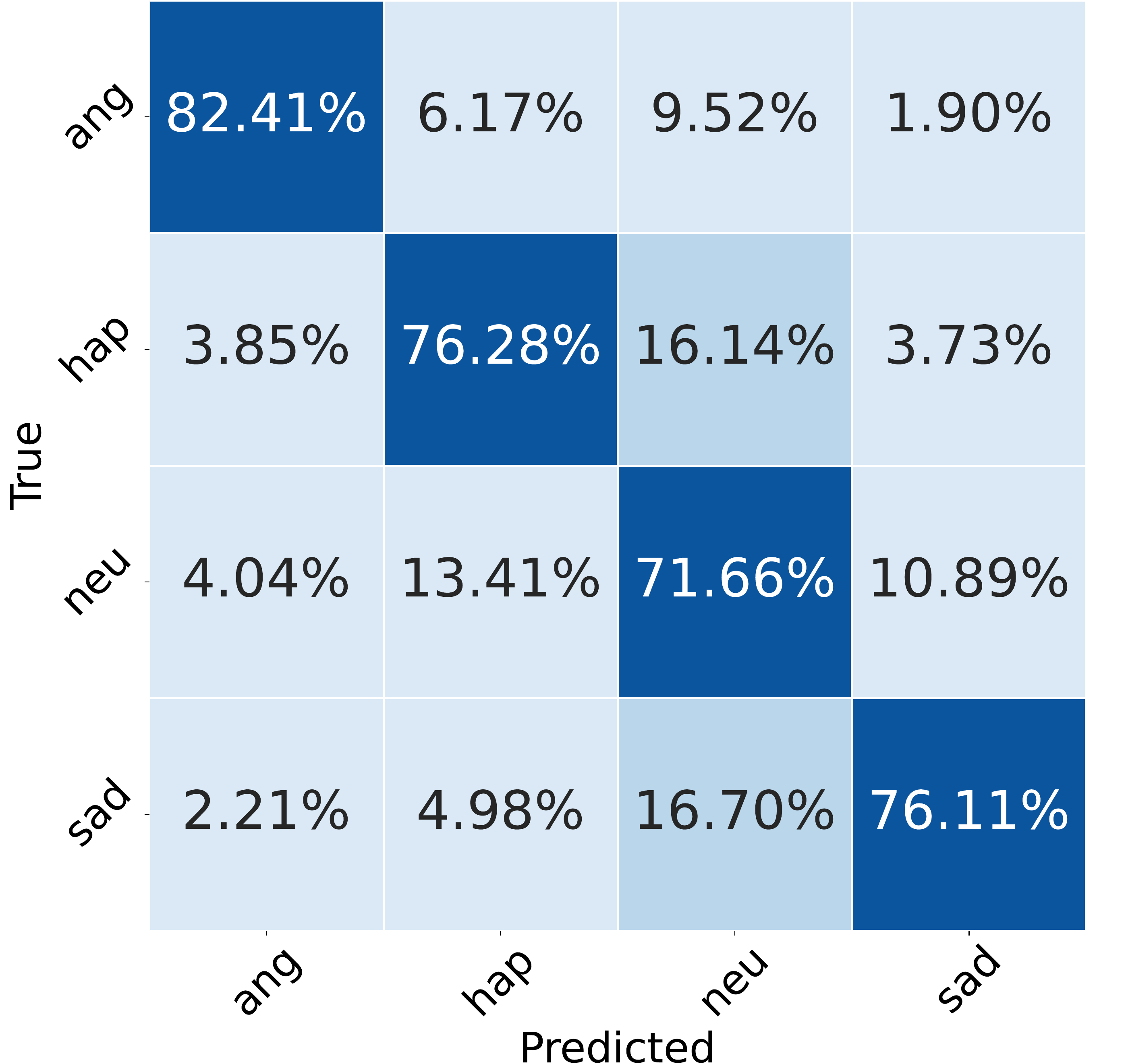}}
  \vspace{0cm}
%  \centerline{(a)}\medskip
\end{minipage}
\hfill
\begin{minipage}[b]{0.48\linewidth}
  \centering
  \centerline{\includegraphics[width=4.6cm]{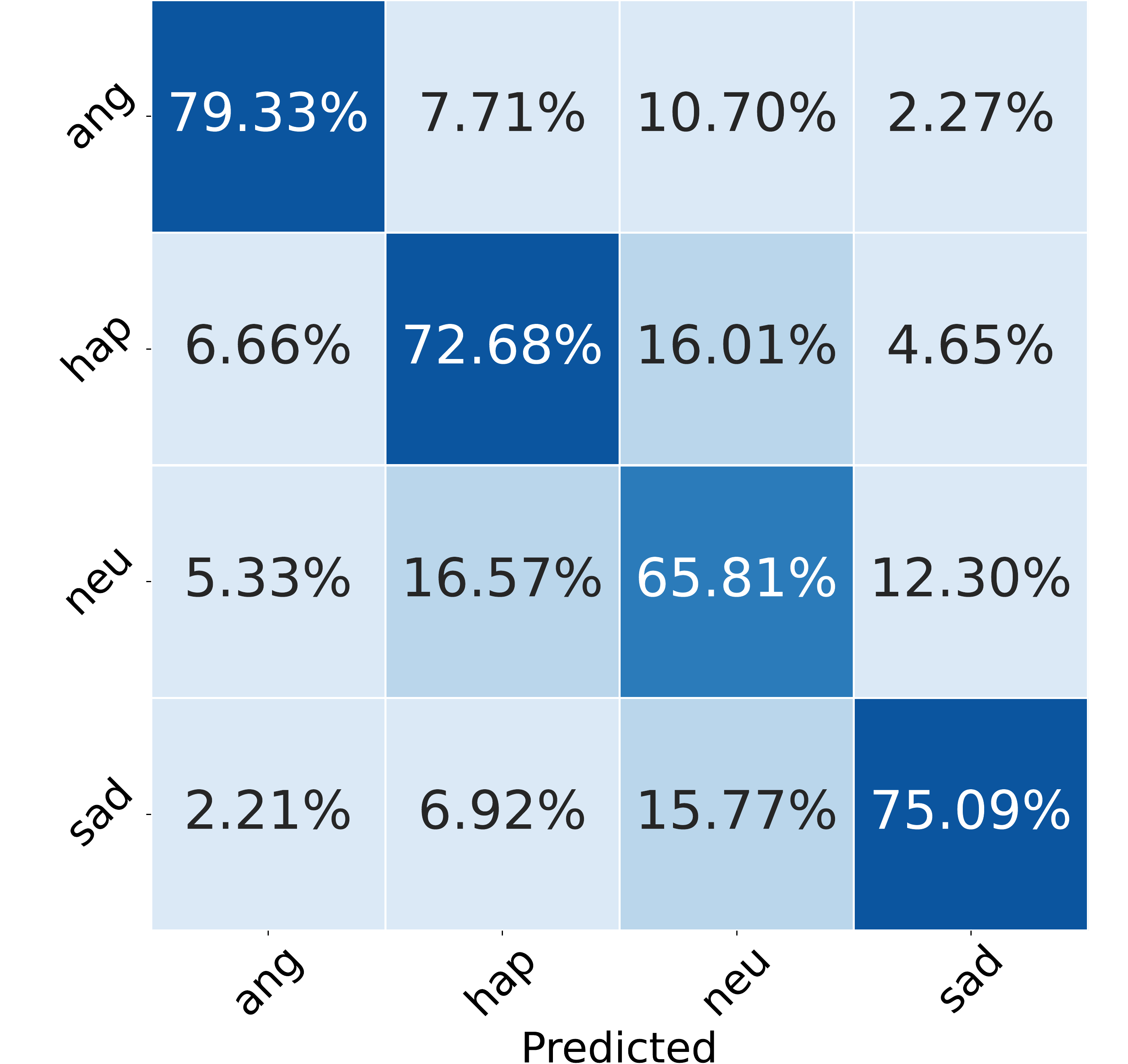}}
  \vspace{0cm}
%  \centerline{(b)}\medskip
\end{minipage}
\hfill
\caption{Confusion matrix for WavLM using attentive correlation (left) and meanstd pooling (right).}
\label{fig:wavlmconfusionmatrices}
\end{figure}

An overview of the results for the most common pooling methods and our proposed approach is shown in Table \ref{tab:emodetectionresults}. The results are presented for the best configuration of our framework with $p_d$ and $p_l$ both equal to 0.25 and $H=4$. The best overall performance was achieved for our proposed approach using WavLM (75.60\%; see also Fig. \ref{fig:wavlmconfusionmatrices}) which is higher compared to other SSL approaches in the literature on the same data --- 67.20\% \cite{pepino2021emotion}, 67.62\% \cite{yang2021superb}, 73.01\% with fine-tuned model \cite{liu2021speech}.

\begin{table}[tbp]
\small %footnotesize, small, normalsize
\centering
\caption{Unweighted accuracy (\% mean and std) between test sets for SER in IEMOCAP using HuBERT large, Wav2vec 2.0, and WavLM large self-supervised representations.}
\label{tab:emodetectionresults}
\begin{tabular}{lccc} %{p{2cm}|p{1.5cm}|p{1.5cm}|p{1.5cm}}
\hline
Pooling method & \textbf{HuBERT} & \textbf{Wav2vec 2.0} & \textbf{WavLM} \\
\hline
 mean & 65.73 (2.73) & 66.86 (1.76) & 69.44 (1.53) \\
 mean-std & 69.15 (1.61) & 69.92 (1.17) & 72.56 (1.67) \\
 corr ($p_d=0$) & 69.82 (1.35) & 68.44 (1.85) & 72.34 (1.54)\\
 corr ($p_d=0.25$) & 69.72 (1.19) & 67.85 (1.84) & 72.27 (1.45)\\
 corr attentive & \textbf{73.86 (2.10)} & \textbf{70.01 (2.20)} & \textbf{75.60 (2.33)}\\
 \hline
\end{tabular}
\vspace{-6mm}
\end{table}

%\begin{figure}[htb]
%\centering
%  \includegraphics[width=\linewidth]{IEMOCAP_WAVLM_emoCorrALSn4Pooling_LabelSmoo%thing0.25_dropout0.25_Confusion_Matrix_5.pdf}
%  \vspace{-8mm}
%  \caption{Confusion matrix for WavLM using attentive pooling.}
%  \label{fig:wavlmattentiveconfusiomatrix}
%  \vspace{-5mm}
%\end{figure}

\begin{figure}[htb]
\centering
  \includegraphics[width=\linewidth]{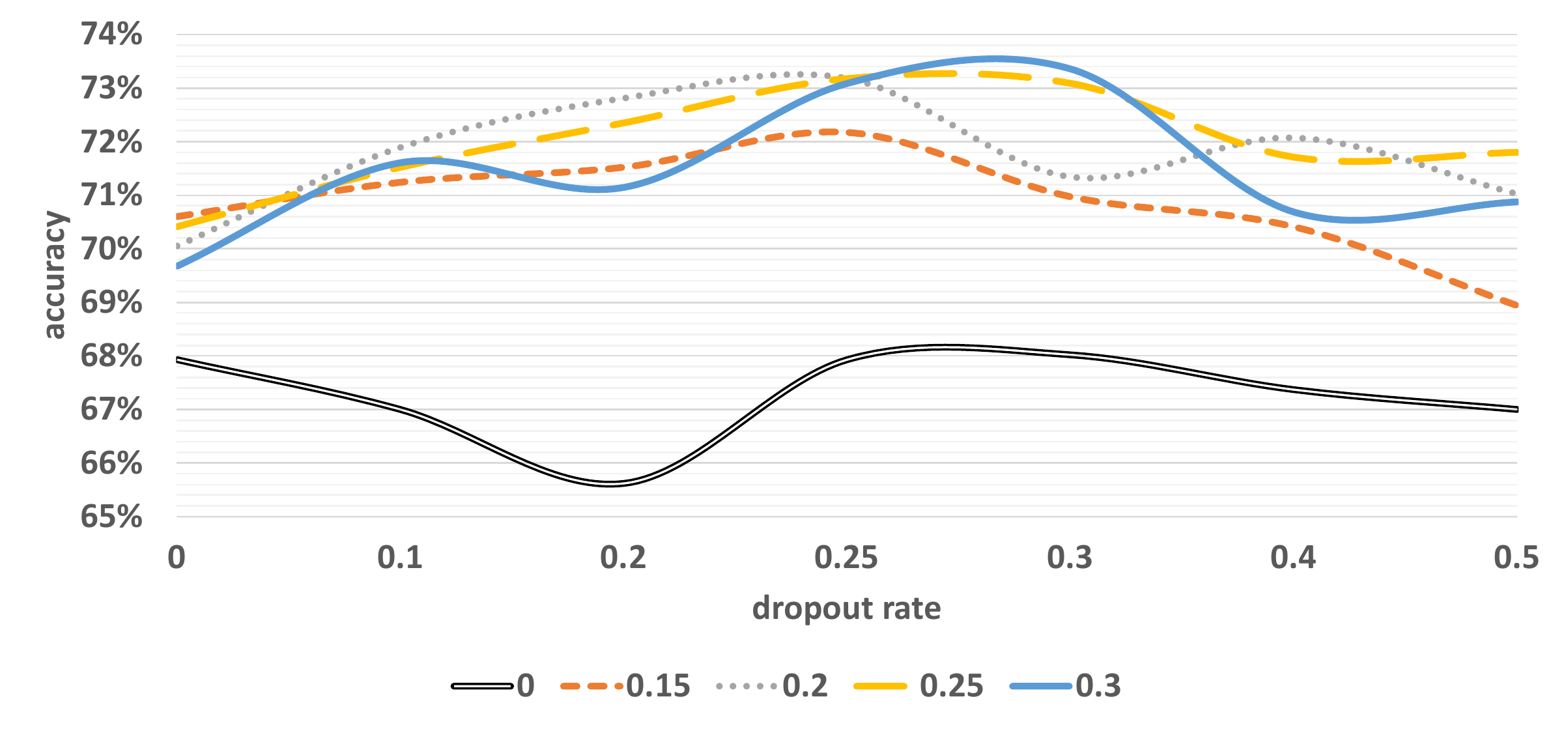}
  \vspace{-8mm}
  \caption{Dropout-label smoothing interaction for Session 1 - HuBERT. Lines represent different label smoothing values.}
  \label{fig:dropoutvslabelsmoothing}
  \vspace{-5mm}
\end{figure}

\subsection{Attention}
\label{ssec:resultsattention}
To investigate the performance of the proposed attentive correlation pooling method we experimented with different numbers of attention heads ($H$). Specifically, we tested for $H = 1, 4, 16, 32$ heads. We obtained the best result with $H=4$. Note that we did not observe any correlation between heads and emotion classes, meaning that there is no direct mapping between heads and emotions. In particular, for Session 1, HuBERT, $p_d=0.25,p_l=0.25$ accuracy was 70.32\% ($H=1$), 73.18\% ($H=4$), 72.53\% ($H=16$), and 71.34\% ($H=32$).

\subsection{Label smoothing and dropout}
\label{ssec:resultssmmothinganddropouts}
To better understand the performance of our system, we probed our setup by varying the parameters for label smoothing and dropout rate. For a set number of attention heads ($H=4$), label smoothing was varied for $p_l=0,0.15,0.25,0.3$ and dropout rate for $p_d=0...0.5$. The effect was evaluated on Session 1 of the setup for attentive correlation pooling using HuBERT and the results can be seen in Fig. \ref{fig:dropoutvslabelsmoothing}. Both dropout and label smoothing have an impact on the performance with label smoothing having a greater positive effect; increasing performance from $67.93\%$ ($p_d=0,p_l=0$) to $70,41\%$ ($p_d=0,p_l=0.25$) and even further with increasing dropout to $73.18\%$ ($p_d=0.25,p_l=0.25$).

The impact of dropout rate and label smoothing was also investigated for mean, mean-std, and correlation pooling. The impact in all was small to negligible. For example, for $p_d=0,p_l=0.25$ the performance remained unchanged compared to $p_d=0,p_l=0$ for mean, mean-std, and correlation pooling while for $p_d=0.25,p_l=0.25$ there was a small improvement for correlation pooling (from $68.20\%$ to $70.60\%$).

\section{CONCLUSIONS}
\label{sec:print}
In this work we presented an SER framework that uses self-supervised representations and is based on label smoothing and a novel approach to attention, attentive correlation pooling. Notably, our method does not require fine-tuning of the pre-trained SSL models but rather uses a light-weight classification head that attempts to capture all relevant emotion information from the pre-trained representations. We run several experiments using a 5-fold cross-validation setup and we have clearly demonstrated that our method reaches high performance in all pre-trained models tested surpassing that of the literature in similar tasks. In future work, we will extend the evaluation setup and validate the performance of our method on more datasets.

% Below is an example of how to insert images. Delete the ``\vspace'' line,
% uncomment the preceding line ``\centerline...'' and replace ``imageX.ps''
% with a suitable PostScript file name.
% -------------------------------------------------------------------------
%\begin{figure}[htb]
%\begin{minipage}[b]{1.0\linewidth}
%  \centering
%  \centerline{\includegraphics[width=8.5cm]{image1}}
%  \vspace{2.0cm}
%  \centerline{(a) Result 1}\medskip
%\end{minipage}
%
%\begin{minipage}[b]{.48\linewidth}
%  \centering
%  \centerline{\includegraphics[width=4.0cm]{image3}}
%  \vspace{1.5cm}
%  \centerline{(b) Results 3}\medskip
%\end{minipage}
%\hfill
%\begin{minipage}[b]{0.48\linewidth}
%  \centering
%  \centerline{\includegraphics[width=4.0cm]{image4}}
%  \vspace{1.5cm}
%  \centerline{(c) Result 4}\medskip
%\end{minipage}
%
%\caption{Example of placing a figure with experimental results.}
%\label{fig:res}
%
%\end{figure}

%\vfill
%\pagebreak

\section{ACKNOWLEDGEMENTS}
\label{sec:acknowledge}
This work was supported by the Academy of Finland project no. 340125 “Computational Modeling of Prosody in Speech”, Czech National Science Foundation (GACR) project NEUREM3 No. 19-26934X, Czech Ministry of Interior project No. VJ01010108 "ROZKAZ" and Horizon 2020 Marie Sklodowska-Curie grant ESPERANTO, No. 101007666. The authors wish to acknowledge CSC – IT Center for Science, Finland, for providing the computational resources. 

%\vfill\pagebreak

\bibliographystyle{IEEEbib}
\bibliography{strings,refs}

\end{document}